# Multi-branch Transmitter for Indoor Visible Light Communication Systems

Safwan Hafeedh Younus[1], Aubida A. Al-Hameed[1,] and Jaafar M. H. Elmirghani[1]
[1]School of Electronic and Electrical Engineering, University of Leeds, LS2 9JT, United Kingdom
elshy@leeds.ac.uk, elaawj@leeds.ac.uk, j.m.h.elmirghani@leeds.ac.uk

**Abstract** - One of the main aims of indoor visible light communication (VLC) systems is to deliver a high data rate service in single user and in multiuser scenarios. A key obstacle is the ability of the indoor VLC channel to support high data rates in the scenarios of interest. Here, we assess the potential of a multi-branch transmitter (MBT) and its use to achieve higher data rates in single user and multiuser indoor VLC systems. For the single user VLC system, the performance of the MBT is examined with a wide field of view (W-FOV) receiver and an angle diversity receiver (ADR) while for the multiuser VLC system we evaluate the performance of the MBT with a non-imaging angle diversity receiver (NI-ADR). In addition, for the multiuser VLC system, we propose subcarrier multiplexing (SCM) tones to allocate an optimum transmitter to each user. Furthermore, wavelength division multiplexing (WDM) is examined to support higher data rates for each user while using on-off-keying (OOK) modulation. In addition, the impact of the user's mobility on the multi-user VLC system performance is studied. The effect of diffuse reflections, mobility and lighting constraints are taken into account. In addition, the effect of co-channel interference (CCI) is considered in the multiuser VLC system.

## I. INTRODUCTION

It is generally accepted that wireless communication has become a very important part of our daily lives. Recent studies by Cisco have shown that data traffic will increase about tenfold by 2020 [1]. It can be seen that the demand for wireless data communication is increasing dramatically. Radio frequencies (RF) are utilized to convey the data. However, due to the increasing demand for wireless data transmission and the fact that most of RF spectrum is occupied, as the congestion in the RF spectrum has increased [2-4]. Consequently, it is beneficial to supplement the RF spectrum to cover the growing demand for wireless data transmission. One of the suggested solutions to overcome the congestion of the RF spectrum is the use of the high-frequency spectrum (beyond 10 GHz) [5]. However, when using this spectrum, some of the favourable propagation properties of RF are lost and the cost of the transceivers increases [2]. In addition, using these bands might lead to an increase in the path loss and an increase in the probability of signal blockages due to the shadowing [5]. It seems that visible light communication (VLC) systems are one of the suitable solutions for dealing with the spectrum crunch in RF systems.

VLC systems have gained attention during the last decade due to the use of light emitting diodes (LEDs) for indoor lighting. It is expected that LEDs will be used to provide 75% of all illumination in the world by 2030 instead of conventional sources of illumination such as fluorescent and incandescent lamps [6]. VLC systems are proposed as complementary systems to RF systems [7] with their huge optical spectrum located between 375 nm and 780 nm [8]. Compared to RF systems, VLC systems offer abundant (hundreds of THz) and license-free bandwidth [9]. In addition, better security is also provided by VLC systems, where light cannot penetrate walls and opaque objects, which means eavesdropping is not possible as in RF systems [10, 11]. Moreover, simple transmitters and receivers (i.e., LEDs and photodetectors) are available at low cost [12].

Several challenges face high data rate VLC systems including the ability of the indoor VLC channel to support high data rates. Transmitters in VLC systems are used for lighting and then for data communication, which means many transmitters with wide beams must be employed to attain the required lighting level in the indoor environments. This results in multipath propagation and limits the indoor VLC channel bandwidth. One of the main challenges in VLC systems is to support multiuser scenarios. The key obstacles include 1) the need to use broad lighting sources to attain an acceptable lighting level in indoor VLC systems, which introduces a very high overlap between the luminaires[13] used (potentially) to serve different users, 2) the interference caused due to the desired and interfering signals [13] received by users, and 3) the change in the shape and the size of the cell in VLC systems, which changes when the direction and level of the illumination provided changes. This can cause the photodetector to treat multiple light sources as one transmitter [9].

One of the main benefits of the indoor VLC system is that the shape, the direction, the intensity and the width of the beam of the luminaires (transmitters) can be controlled to achieve an acceptable level of illumination in the indoor environment [14] while also supporting optimum communication. By controlling one or all of these parameters of the luminaires, the channel of the indoor VLC system can be improved to support higher data rates and support multi-user scenarios. Recent research has shown that when using computer-generated holograms (CGHs), the beams of the light units can be directed to improve the 3dB channel bandwidth and to boost the performance of the VLC system [15], [16]. However, using CGHs increases the complexity of the VLC system. Thus, we use a multi-branch transmitter (MBT) as a solution to improve the indoor VLC channel's properties and support multi-user communication. The MBT has many transmitter branches (TBs) and each one is directed to a specific area. Due to the reduction in the semi-angle of each TB, the effect of multipath propagations is reduced and the received optical power is improved.

Multi beam transmitters have been studied in indoor VLC systems. Space division multiple access (SDMA) was realised by using MBTs to serve many users simultaneously [17], [18]. It was shown that increasing the number of TBs improves the performance of the VLC system. However, the effect of the diffuse reflections was not considered in [17], [18]. In [19], the MBT was used to split the communication area into small sections (attocells) to mitigate the interference between users. However, perfect channel knowledge was supposed between the transmitter and the receiver in [19]. Due to the directivity of the MBT, it was used to estimate accurately the position of the receiver [20], [21]. The MBT was also used to assign a group of LEDs for users [22]. It was shown that using multi-element receivers with MBT can improve the performance of the VLC system [22]. However, the effect of mobility on the performance of this system was not considered [22].

It should be noted that the data rates achieved by [17]-[22] are still low when compared to the available VLC spectrum. In addition, [17]-[22] assumed that the locations of the receivers are known for resource allocation. However, the location information of the users may not be available. Thus, in this work, we focused on two main challenges in multi-user indoor VLC systems. These are i) designing a multi-user VLC system that achieves high data rates and ii) providing a new resource allocation method for indoor multi-user VLC systems that does not call for knowledge of the receiver location. We used a MBT in conjunction with a multi-colour LD, which enables wavelength division multiplexing (WDM) and consequently increased the data rate of each user. In addition, using WDM enables each branch of the MBT to serve up to four users by allocating a different channel for each user. For the resource allocation problem, we proposed subcarrier multiplexing (SCM) tones where the user is assigned its best transmitter without needing to know the user (receiver) location. To the best of our knowledge, the data rates achieved by our proposed system are the highest data rates reported in a multi-user indoor VLC system. In addition, this is the first time that SCM tones have been used as a resource allocation tool in multi-user indoor VLC systems. It is worth noting that in our design we considered the effect of the azimuth and the elevation of the MBT on the coverage area of each branch of the MBT. We provide a mathematical model for the MBT in which the effect of the elevation and azimuth are taken into account to obtain the irradiance angle of each face for the MBT.

We use the MBT to enhance the performance of the VLC systems for the single user and multiuser scenarios while considering the effects of diffuse reflections (up to second order reflections), acceptable illumination level in the environment, mobility and co-channel interference (CCI) between luminaires (for the multiuser scenario). We first evaluate the performance of the MBT with a single user VLC system. In this case, we obtain the delay spread, 3dB channel bandwidth and signal to noise ratio (SNR) of the user, which are the important factors that measure the performance of the VLC system. We evaluate the performance of this system with two types of receivers: wide field of view (FOV) receiver and angle diversity receiver (ADR). The results show that this system can provide a data rate of 4 Gb/s and 10 Gb/s when using a wide FOV receiver and an ADR, respectively.

Secondly, we use the MBT with wavelength division multiplexing (WDM) and SCM tones to realise a high data rate multiuser indoor VLC system. Each TB can be used to send a different data stream; therefore, many users can be served simultaneously. Here, we used SCM tones to i) find the optimum TB for each optical receiver and ii) determine the level of the CCI between luminaires. In addition, these SCM tones might help with the handover operations during user mobility. Four laser diodes (RYGB LDs) are used in this work as indoor lighting sources as well as modulators. To set up the link between the transmitters and receivers, one colour of RYGB LDs are utilized to send the SCM tones at the start of the communication session. When the connection is set up, the data is transmitted in parallel through the RYGB LDs. We use WDM to improve the data rate for each user. The multiuser VLC system is investigated with an array of non-imaging angle diversity receivers (NI-ADR).

The rest of this paper is organised as follows: Section II describes the simulation's setup. Section III describes the structure of the MBT. Evaluation of the performance of the MBT with a single user VLC system is given in Section IV. The MBT for the multiuser VLC system is described in Section V. The structure of the NI-ADR for the multiuser scenario is given in Section VI. Section VII shows the performance of the multiuser VLC system. Section VIII shows the effect of the people mobility on the system performance. Finally, conclusions are given in Section IX.

## II. SIMULATION SET-UP

A simulation was developed in an unfurnished room, which does not have doors and windows, to evaluate the performance of our proposed systems. The room has a length of 8 m, a width of 4 m and a height of 3 m. Previous work has shown that the pattern of the reflected light rays from plaster walls is approximately Lambertian [23, 24]. Hence, room reflecting surfaces are modelled as Lambertian reflectors. The room's ceiling and walls have reflection coefficients of 0.8 while the room's floor has a reflection coefficient of 0.3 [25]. To model reflections from the room's surfaces, we used a ray tracing method. Consequently, we divided the room's surfaces into a number of small surface elements. These surface elements are equal sized, square-shaped and have an area of $d_A$ with reflection coefficients of $\rho$ as shown in Fig. 1. These surface elements were assumed as secondary small emitters, which reflect the received optical signals in the shape of a Lambertain pattern with $n = 1$, where $n$ is the order of the emission of the Lambertain beam. We considered reflections up to the second order. It should be noted that results with higher resolutions can be obtained if the surface element's size is reduced. Reducing the size of the surface elements increases the computation time. Thus, the size of the surface elements for the first order reflections was set as 5 cm × 5 cm while it was set as 20 cm × 20 cm for second order reflections. This completes the computations within a moderate time [25].

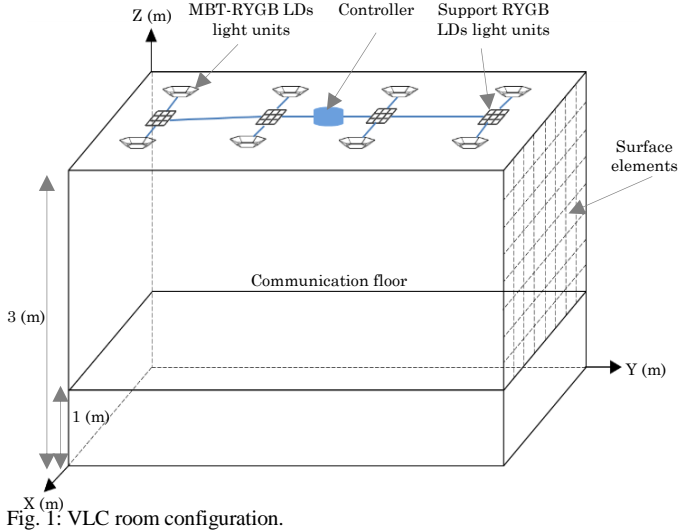

Fig. 1: VLC room configuration.

In this work, RYGB LDs were used instead of LEDs as luminaires. This was mainly due to the wider modulation bandwidth of the LDs compared with LEDs. Previous research has concluded that RYGB LDs with a diffuser can be used as illumination sources without any risk to the human eye [26, 27]. In the simulation, all luminaires were located on the room's ceiling (3 m above the floor). As we used a diffuser, the RYGB LDs emission pattern is considered Lambertian. The illumination level associated with LOS links and reflection links (up to second order) was calculated as in [28], [29], [30].

In the VLC system, intensity modulation / direct detection (IM/DD) is the simplest format of modulation [31]. Propagation in the multipath channel when using IM/DD can be fully modelled by the channel impulse response ($h_k(t)$) [32, 33]:

$$I(t, Az, El) = \sum_{k=1}^{L} R\, x(t) \otimes h_k(t, Az, El) + \sum_{k=1}^{L} R\, n_k(t, Az, El) \quad (1)$$

where $I(t, Az, El)$ is the instantaneous current received by the photodetector, $t$ is the absolute time, $Az$ and $El$ are the arrival directions, $L$ denotes the total number of receiving elements, $R$ is the responsivity of the photodetector, $x(t)$ is the instantaneous optical power transmitted by the transmitter, $\otimes$ denotes convolution and $n(t, Az, El)$ is the background received noise. Using numerical simulation [25, 34, 35], the impulse response can be evaluated and several parameters can be obtained. These include the power distribution, the SNR and root-mean-square (rms) delay spread (D). Due to diffuse reflections, the indoor VLC systems are subject to multipath dispersion, which results in inter symbol-interference (ISI). The delay spread is given as [36]:

$$D = \sqrt{\frac{\sum (t_i - \mu)^2 P_{ri}^2}{\sum P_{ri}^2}} \quad (2)$$

where $t_i$, $P_{ri}$ and $\mu$ are the delay time of a ray, received optical power and the mean delay, respectively. The mean delay, $\mu$, is given by:

$$\mu = \frac{\sum t_i P_{ri}^2}{\sum P_{ri}^2} \quad (3)$$

### III. MBT Structure

The MBT is a group of TBs in which each TB is oriented to a different direction and covers a small different part of the room, as shown in Fig. 2. In this work, the MBT has seven TBs (1 to 7), and each TB has two white RYGB LDs with narrow-semi angles. The two RYGB LDs were used in each branch to give illumination at an acceptable level in the room, which meets the required level of lighting. The coverage area of each TB can be modified by changing the Lambertian emission order ($n$). The value of $n$ should be selected to reduce the overlap between the TBs and to keep the illumination level at the desired value. In this work, the Lambertion emission order of each RYGB LD in the TB was 11, which gives a semi angle equal to 20.1°. This leads to overlap between adjacent TBs of up to 5.6% and provides an acceptable illumination as shown in Fig. 2. It should be noted that there is a trade-off between the illumination level and the overlap percentage. If the percentage of the overlap between the two adjacent TBs is zero, this leads to gaps between branches and reduces the illumination level in some areas to a level under the recommended value (i.e., 300 lx [37]). On the other hand, a decrease in the value of $n$ leads to an increase in the overlap between the TBs that consequently increases the interference.

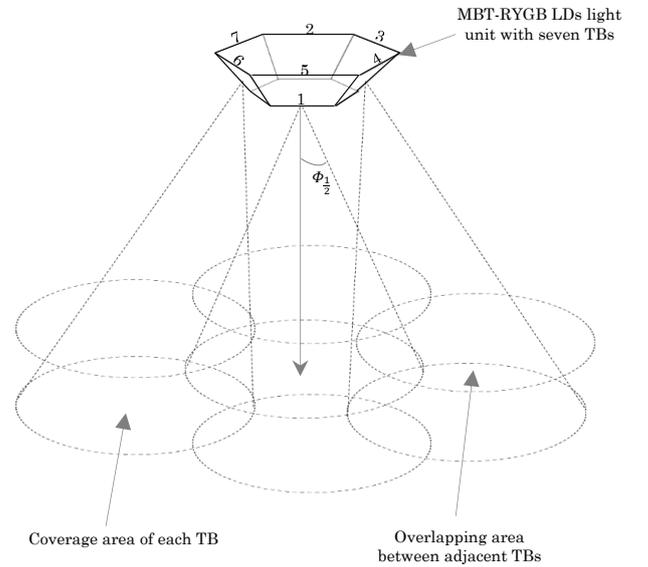

Fig. 2: Structure of MBT.

Each branch in the MBT has a certain orientation that is defined by two angles: azimuth ($A_z$) and elevation ($El$). In this work, the $El$ angle of the first TB was set at 90°, and the other six TBs were given an $El$ of 60°. The $A_z$ angles of the seven TBs were fixed at 0°, 0°, 60°, 120°, 180°, 240° and 300°. The values of the $Els$ and $A_{zs}$ of the MBT were optimised to give sufficient illumination and to realize good link quality between the transmitters and receivers [38] at all possible places in the room's communication floor.

To compute the irradiance angle ($\theta$) for any TB, the $A_z$ and $El$ should be taken into account. Therfore, a point $P$ was defined and located on the transmitter's normal, 1 m under the transmitter, as shown in Fig. 3. This point is located in the FOV of the TB and is assumed to have a 1 m distance from the TB to ease the analysis. Fig. 3 also depicts the transmitted light from a TB to the receiver. The angle $\theta$ can be given as:

$$\theta = cos^{-1}\left(\frac{|\overrightarrow{PT_x}|^2 + |\overrightarrow{R_xT_x}|^2 - |\overrightarrow{PR_x}|^2}{2\,|\overrightarrow{PT_x}|\,|\overrightarrow{R_xT_x}|}\right) \quad (4)$$

where:

$$|\overrightarrow{R_xT_x}|^2 = (x_r - x_t)^2 + (y_r - y_t)^2 + (z_r - z_t)^2 \quad (5)$$

$$|\overrightarrow{PT_x}|^2 = 1 + d^2 \quad (6)$$

and

$$|\overrightarrow{PR_x}|^2 = (x_p - x_r)^2 + (y_p - y_r)^2 + (z_p - z_r)^2 \quad (7)$$

From Fig. 2, we can also calculate the:

$$El = tan^{-1}\left(\frac{1}{d}\right) \quad (8)$$

$$x_p = x_t + \frac{cos(A_z)}{tan(El)} \quad (9)$$

$$y_p = y_t + \frac{sin(A_z)}{tan(El)} \quad (10)$$

and

$$z_p = z_t - 1 \quad (11)$$

By substituting (9) – (11) in (7), $PR_x$ can be rewritten as:

$$|\overrightarrow{PR_x}|^2 = \left(\left(x_t + \frac{cos(A_z)}{tan(El)}\right) - x_r\right)^2 + \left(\left(y_t + \frac{sin(A_z)}{tan(El)}\right) - y_r\right)^2 + \left((z_t - 1) - z_r\right)^2 \quad (12)$$

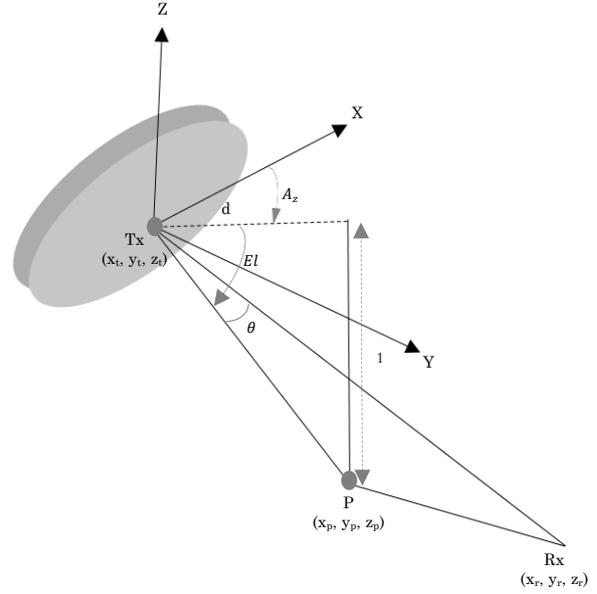

Fig. 3: Elevation and azimuth analysis for MBT.

Eight MBT-RYGB LD light units are fitted on the room's ceiling and utilized for lighting and communication. Each MBT-RYGB LDs light unit has seven branches and each branch has two RYGB LDs. However, due to the narrow FOV of each branch in the MBT, the lighting illumination level recommended by the ISO and European standards [37] cannot be maintained (i.e., illumination will be lower than 300 lx in some room locations), as can be seen in Fig. 4 (a). Therefore, additional RYGB LD light units (four support RYGB LD light units) were added to enhance the lighting level (see Fig. 1). These support RYGB LD light units that are utilized for lighting only and each unit has $3 \times 3$ RYGB LDs. Hence, the required illumination level in the room was achieved as shown in Fig. 4 (b). It should be noted that an increase in the number of MBT-RYGB LD light units and/or the number of RYGB LDs in each branch of the MBT-RYGB LDs light units could not help achieve the target lighting level (i.e., 300 lx). Therefore, the support RYGB LD light units were added to enhance the illumination level (these support light units can be either LDs or LEDs as they are not used for communication).

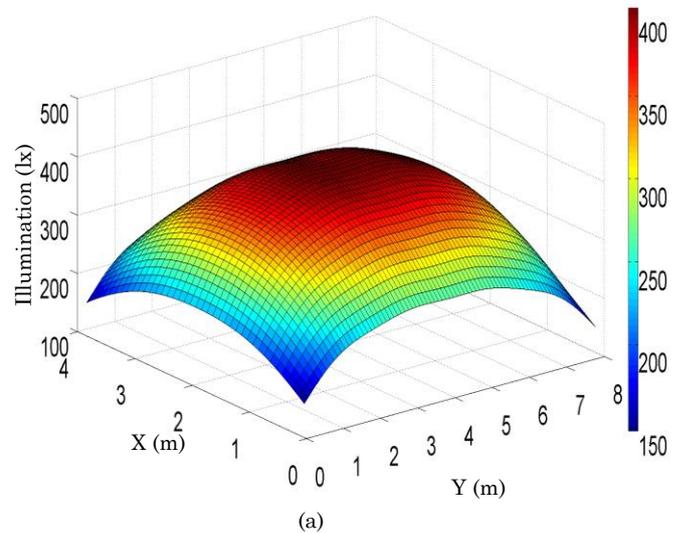

(a)

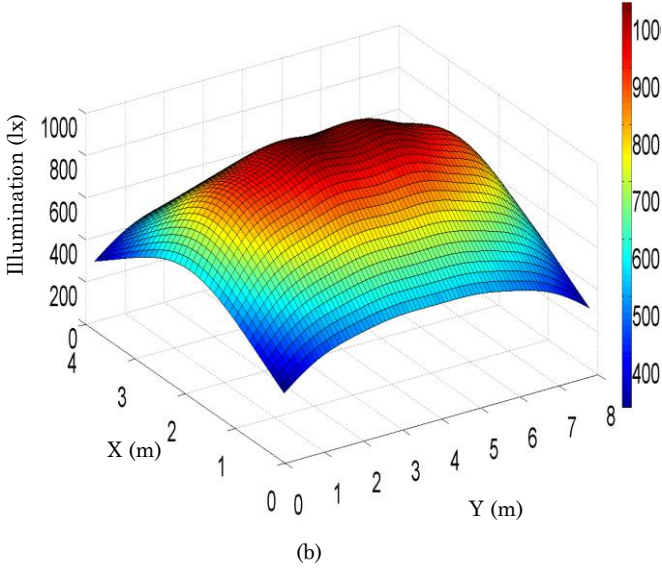

(b)

Fig. 4: Distribution of lighting in room: (a) without support of RYGB LD light units (min. illumination 107 lx and max. illumination 403 lx) and (b) with support of RYGB LD light units (min. illumination 305 lx and max. illumination 1012 lx).

The room and light unit parameters are shown in Table I.

TABLE I
SIMULATION PARAMETERS

| Parameters | Configurations | |
|---|---|---|
| **Room** | | |
| Length | 8 m | |
| Width | 4 m | |
| Height | 3 m | |
| $\rho$-xz Wall | 0.8 | |
| $\rho$-yz Wall | 0.8 | |
| $\rho$-xz op. Wall | 0.8 | |
| $\rho$-yz op. Wall | 0.8 | |
| $\rho$-Floor | 0.3 | |
| Bounces | 1 | 2 |
| Surface elements no. | 32000 | 2000 |
| $d_A$ | 5 cm × 5 cm | 20 cm × 20 cm |
| Emission order of Lambertian ($n$) | 1 | |
| **Support RYGB LDs light units** | | |
| Number of units | 4 | |
| Locations (x, y, z) m | (2, 1, 3), (2, 3, 3), (2, 5, 3), (2, 7, 3) | |
| Number of RYGB-LDs/ unit | 9 (3 × 3) | |
| Elevation | 90° | |
| Azimuth | 0° | |
| Transmitted optical power/RYGB LDs | 1.9 W | |
| Centre luminous intensity/RYGB LDs | 162 cd | |
| Emission order of Lambertian ($n$) | 0.65 | |
| **MBT- RYGB LDs light units** | | |
| Number of units | 8 | |
| Number of TBs/unit | 7 | |
| Locations(x, y, z) m | (1, 1, 3), (1, 3, 3), (1, 5, 3), (1, 7, 3), (3, 1, 3), (3, 3, 3), (3, 5, 3), (3, 7, 3) | |
| Elevation/TB | 90°, 60°, 60°, 60°, 60°, 60°, 60° | |
| Azimuth/TB | 0°, 0°, 60°, 120°, 180°, 240° 300° | |
| Transmitted optical power/RYGB LDs | 1.9 W | |
| Centre luminous intensity/RYGB LDs | 162 cd | |
| Lambertian emission order ($n$) | 11 | |

## IV. SINGLE USER MBT VLC SYSTEM PERFORMANCE.

This section reports the performance of the MBT for the single user VLC system. Two receivers were used in this section: a wide field of view (W-FOV) receiver and an angle diversity receiver (ADR). The W-FOV receiver has a responsivity of 0.4 A/W and FOV of 40° to enable it to view at least one transmitter at any place on the room's communication floor. In addition, the detection area of the W-FOV receiver was selected to be 1 mm², which enables the W-FOV receiver to work at a high data rate up to 4.4 Gb/s [15]. The ADR is a number of photodetectors and these photodetectors have narrow FOVs and are directed in different directions. In this work, the ADR consisted of seven detector faces (1-7) with photodetectors that have a responsivity of 0.4 A/W. Two angles were used to define the direction of each branch in the ADR: $A_z$ and $El$. The $El$ angle of the first detector was set at 90°, whereas the other six detectors were set up at an $El$ of 60°. The azimuth ($A_z$) angle is the direction of the detector's angle, and the $A_z$ angles of the seven detectors were selected to be 0°, 0°, 60°, 120°, 180°, 240° and 300°. Each face of the ADR has a FOV of 20°. In the ADR, each photodetector has an active area 0.4 mm². The optical power received by each face of the ADR can be amplified separately. Hence, diverse methods can be used to combine the signals, for example: maximum ratio combining (MRC), select the best (SB) and equal gain combining (EGC). In the single user VLC system, we used the SB scheme to obtain the results. The W-FOV receiver and each photodetector in the ADR employed a compound parabolic concentrator (CPC) [31]. This CPC has an acceptance angle ($\psi$) of less than 90° and a gain ($g(\psi)$) given by [31]:

$$g(\psi) = \begin{cases} \dfrac{N^2}{\sin^2\psi}, & 0 \leq \Upsilon \leq \psi \\ 0, & \Upsilon \geq \psi \end{cases} \quad (13)$$

where $N$ and $\Upsilon$ denote the refractive index and the angle of the incidence, respectively.

A simulation tool similar to that utilized by Barry *et al* [39] was developed and used here to obtain MBT single user VLC system results. The simulation tool is utilized to obtain the impulse responses and to calculate the delay spread and SNR. The effect of mobility and reflections on the VLC system performance were taken into account. To consider the effect of mobility, we obtained the results when the mobile user moves along $x = 0.5$ m and along $x = 2$ m, which represent the worst-cases as the ISI is high at $x = 0.5$ m and the distance between the MBT-RYGB LD light units and the mobile user is at a maximum at $x = 2$ m. At each location of the mobile user, one TB is allocated to the optical receiver. A "select the best TB" algorithm is utilized to assign the mobile user to the best TB. The operation of the "select the best TB" algorithm can be summarized as follows:

1. The controller allocates an ID to each TB.
2. The TBs are turned ON individually by the controller.
3. The SNR of each TB is obtained by the receiver.
4. The receiver sends a low data rate infrared (IR) feedback signal to notify the controller of the SNR attributed to each TB. We used the IR uplink design in [40].

5. The TB that gives the highest SNR is considered by the controller to send data and the other TBs are utilized for illumination only (no data is transmitted through these TBs).

Due to the symmetry of the room, more than one TB can provide the same SNR in some locations of the optical receiver. In this case, the controller considers one TB and ignores the other TBs. In addition, the transmitted data is modulated by the two RYGB LDs in each TB for the single user VLC system.

*A. Impulse response*

The impulse responses of the W-FOV receiver and the ADR when the mobile user was at (0.5 m, 0.5 m, 1 m), which represents the worst-case as the ISI is high at this location, are shown in Fig. 5 (a) and Fig. 5 (b), respectively. Despite the fact that the detection area of the W-FOV receiver is larger than the detection area of the ADR, and the LOS optical power at the ADR is higher when compared to the LOS optical power of the W-FOV receiver. This is due to the difference in the gain of the CPCs that were used in the ADR and W-FOV receiver. In addition, the impulse response of the ADR (see Fig. 5 (b)) is better than the impulse response of the W-FOV receiver (see Fig. 5 (a) and Fig. 5(b)) in terms of signal spread, which decreased the delay spread and improved the 3dB channel bandwidth. This was due to the narrow FOV for each branch of the ADR compared to the FOV of the W-FOV receiver, which limited the number of rays captured by the ADR.

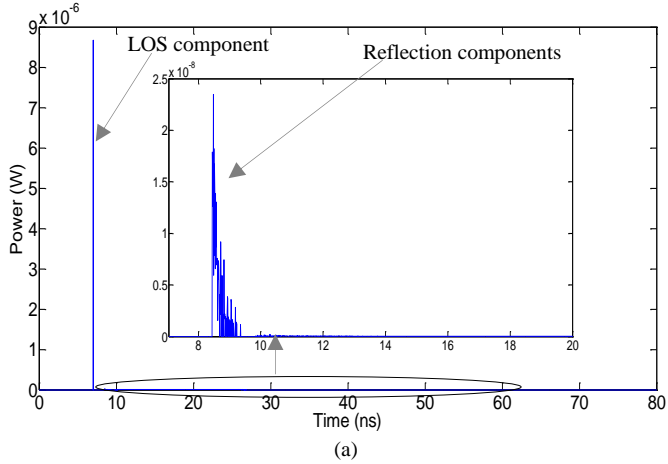

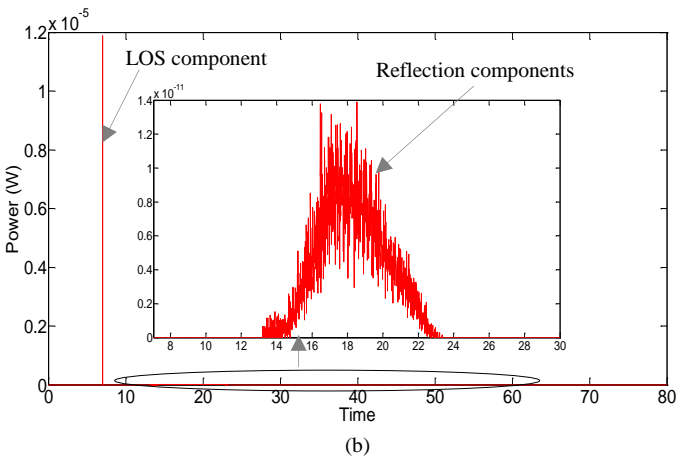

Fig. 5: Impulse responses when mobile user is located at (0.5 m, 0.5 m, 1 m): (a) W-FOV receiver and (b) ADR.

*B. Delay spread evaluation and 3dB channel bandwidth*

Fig. 6 illustrates the delay spread of the W-FOV receiver and the ADR when the mobile user moves along the *y*-axis on the communication floor at $x = 0.5$ m and at $x = 2$ m. The received optical power of each ray impacts the delay spread; therefore, a reduction in the collected power from the reflection components leads to a decrease in the delay spread. Thus, as can be seen, the delay spread of the W-FOV receiver is worse than the delay spread of the ADR along $x = 0.5$ m. This is attributed to the large FOV of the W-FOV receiver, which means that the number of rays that were captured by the W-FOV receiver is larger when compared with the ADR. When the mobile user moves at $x = 2$ m, the ADR offers better performance over the W-FOV at (2 m, 0.5 m, 1 m) and (2 m, 7.5 m, 1 m) only. This is due to the locations of the W-FOV receiver and the ADR which were located far from walls at $x = 2$ m and the TB that served the mobile user at $x = 2$ m directed to the communication floor of the room, which means reflections from the walls and the ceiling of the room are very low. Thus, the performance of the W-FOV is comparable with that of the ADR. We can conclude that the ADR offers better performance over the W-FOV receiver when the optical receiver is placed close to the walls of the room.

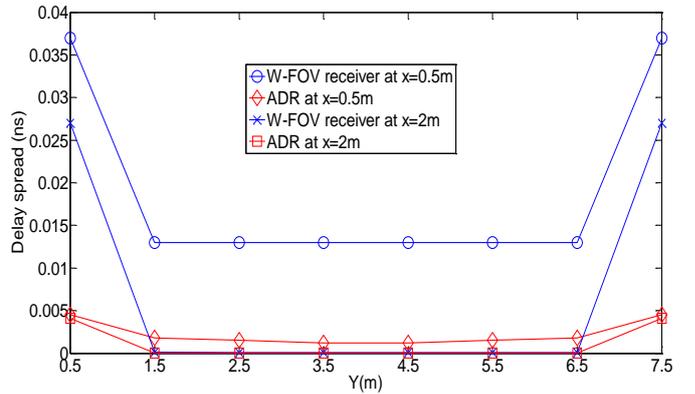

Fig. 6: Delay spread of W-FOV receiver and ADR when mobile user moves along *y*–axis and at $x = 0.5$ m and $x = 2$ m.

Table II shows the 3dB channel bandwidth when the two receivers (W-FOV receiver and ADR) are used. The results were obtained when the mobile user moves along the *y*-axis and at $x = 0.5$ m and $x = 2$ m. In general, the ADR provides bandwidth larger than that of the W-FOV receiver. This is due to the reduction in the effect of the diffuse reflections in ADR due to its narrow FOV of each photodetector, which significantly reduces the delay spread and increases the bandwidth. It can be seen that the lowest value of the 3dB channel bandwidth is 4.5 GHz for the W-FOV receiver and 22.3 GHz for the ADR. The 3dB channel bandwidth enables the VLC systems to support data rates up to 6.4 Gb/s for W-FOV receiver and up to 31.8 Gb/s for ADR without ISI while utilizing OOK modulation where an optimum receiver bandwidth (in optical direct detection systems) of 0.7 times the bit rate is assumed [41]. When the optical receiver is placed in the middle of the room (at $x = 2$ m), the indoor VLC system can be assumed to have a flat channel for both receivers due to the low reflection components when the optical receiver was moved at $x = 2$ m.

TABLE II
MOBILE USER 3dB CHANNEL BANDWIDTH ALONG Y-AXIS

| Y(m) | 3dB Channel bandwidth (GHz) | | | |
|---|---|---|---|---|
| | Receiver at x = 0.5 m | | Receiver at x = 2 m | |
| | W-FOV | ADR | W-FOV | ADR |
| 0.5 | 4.5 | 22.3 | 6.17 | 24.4 |
| 1.5 | 7.7 | 55.6 | Flat channel | Flat channel |
| 2.5 | 7.7 | 56.1 | Flat channel | Flat channel |
| 3.5 | 7.7 | 56 | Flat channel | Flat channel |
| 4.5 | 7.7 | 56 | Flat channel | Flat channel |
| 5.5 | 7.7 | 56 | Flat channel | Flat channel |
| 6.5 | 7.7 | 56 | Flat channel | Flat channel |
| 7.5 | 4.5 | 22.3 | 6.17 | 22.4 |

*C. SNR Evaluation*

The performance of the MBT for single user indoor VLC system can be strongly impaired by mobility, ISI and multipath propagation. The simplest modulation technique for the indoor VLC system is the OOK, and the BER of the conventional OOK modulation technique for the indoor VLC system is written as [31]:

$$BER = \frac{1}{2} \, erfc(\sqrt{SNR/2}) \quad (14)$$

where $erfc$ is the complementary error function. The $SNR$ is given as [42]:

$$SNR = \left(\frac{R(P_{s1} - P_{s0})}{\sigma t_t}\right)^2 \quad (15)$$

here $\sigma_t$ is the standard deviation of the total noise and can be written as [43]:

$$\sigma_t = \sqrt{\sigma_{bn}^2 + \sigma_s^2 + \sigma_{pr}^2} \quad (16)$$

where $\sigma_{bn}$ is the ambient shot noise, $\sigma_s$ denotes the shot noise related to the data signal and $\sigma_{pr}$ is the pre-amplifier thermal noise. The background light shot niose ($\sigma_{bn}$) is given as:

$$\sigma_{bn} = \sqrt{2qAI_{bn}BW} \quad (17)$$

where $q$ is the electron charge, $I_{bn}$ is the background photocurrent per unit area (of the photodetector), which is induced due to the light from the sky and background light sources ($I_{bn} = 10^{-3}$ A/cm$^2$) [23], $A$ is the photodetector area and $BW$ is the bandwidth of the pre-amplifier, The shot noise induced by the data signal is expressed as [44]:

$$\sigma_s = \sqrt{2qRPr_nBW} \quad (18)$$

In this paper, we used the p-i-n FET receiver design in [45], which has an input noise current equal to 4.5 pA/$\sqrt{Hz}$.

Fig. 7 (a) shows the SNR results of the two receivers at a bit rate of 4 Gb/s. The SNR was obtained when the mobile moves along the y-axis and $x = 0.5$ m and $x = 2$ m. The lowest value of the SNR achieved by the W-FOV receiver was 13.2 dB (see Fig. 7 (a)) when the mobile user was placed at (0.5 m, 0.5 m, 1 m). This means that the W-FOV receiver provides a good connection between the receiver and transmitters at a data rate of 4 Gb/s. A significant enhancement was achieved in the SNR at a data rate of 4 Gb/s when the ADR was utilized instead of the W-FOV receiver along $x = 0.5$ m. This improvement in the SNR is due to the narrow FOV of each face of the ADR, which reduces the reflection components. However, the performance of the W-FOV was comparable to that of the ADR when the optical receivers were placed at the room's centre (i.e., along $x = 2$ m) as shown in Fig. 7 (a). This is because of the placement far from the walls of the room at $x = 2$ m. In addition, the TB that served the mobile user at $x = 2$ m is directed to the communication floor of the room, which means reflections from walls and the ceiling of the room are very low. Thus, both receivers can offer good performance at a data rate of 4 Gb/s in the centre of the room.

Fig. 7 (b) illustrates the mobile user SNR of the ADR when the VLC system works at 10 Gb/s. We only show the results of the ADR at a data rate of 10 Gb/s. This is due to the 3dB channel bandwidth produced by the W-FOV receiver (see Table II) which is not able to support 10 Gb/s when the optical receiver was placed at the room's corner (at (0.5 m, 0.5 m, 1m ) and at (0.5 m, 7.5 m, 1 m)). In addition, the photodetector area of the W-FOV receiver is 1 mm$^2$, which allows it to work at a data rate up 4.4 Gb/s. We emphasise that the performance of the W-FOV receiver is comparable to the ADR when the optical receiver is placed far from the walls of the room where the reflection components are very low. It can be seen that the lowest SNR is 14.2 dB (when the ADR worked at a data rate of 10 Gb/s), which gives a BER of $1.5 \times 10^{-7}$.

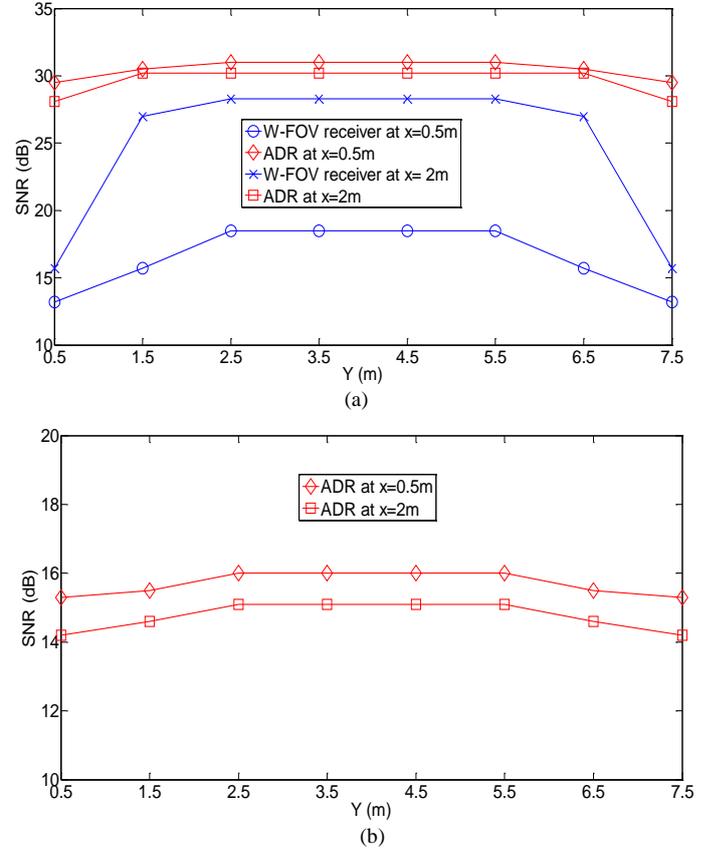

Fig. 7: W-FOV receiver and ADR SNR (a) SNR of the two receivers working at 4 Gb/s and (b) SNR of ADR operating at 10 Gb/s, at $x = 0.5$ m and at $x = 2$ m along the *y*-axis.

## V. MBT FOR MULTIUSER INDOOR VLC SYSTEMS.

As each TB is directed to a (relatively large) specific location on the room's communication floor, many users can be served simultaneously. In addition, we used RYGB LDs as luminaires; hence, WDM can be used to enable each TB to serve up to four users at the same time. Allocating an optimum transmitter to each user in a multiuser indoor VLC system is one of the main challenges that need to be tackled. In this section, we propose the use of SCM tones to assign a TB to each user. SCM tones were studied in a range of applications in indoor optical wireless communication systems [46-49]. The work in [46] proposed SCM tones to realise a high data rate indoor VLC system using parallel data transmission. The SCM tones were utilized to determine the CCI level and also to match transmitters with the imaging receiver pixel(s). The level of the crosstalk between the channels of WDM was calculated using SCM tones in [47]. In [48], SCM tones were proposed for an indoor VLC positioning system. The SCM tones were used also to allocate an optimum transmitter to each user in a multiuser VLC system [49].

Each user is served by the TB that offers the best in terms of received optical power and low CCI level. These SCM tones are unmodulated tones and were proposed to recognise each TB, allocate the best TB to each user, calculate the CCI between the TBs and can be used to managing the handover during user mobility. Any colour of the RYGB LDs can be used to send the SCM tones. We used the green colour to send the SCM tones at the beginning of the transmission for setting up the communication between TBs and users. Once the link between the transmitters and optical receivers is set up, data is transmitted through all four colours of the RYGB LDs. Fig. 8 shows the structure of the RYGB LDs for the multiuser scenario.

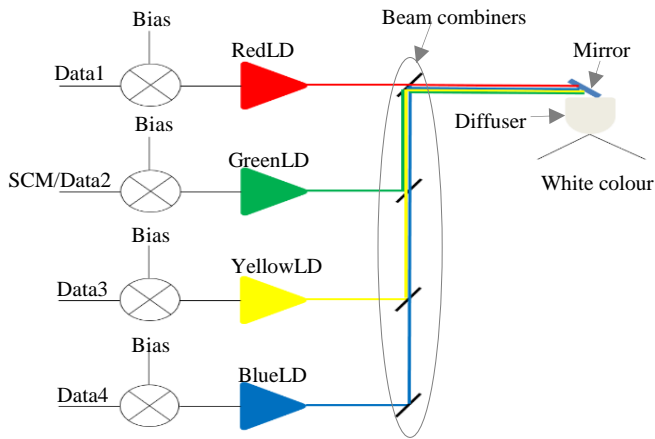

Fig. 8: RYGB LDs structure.

## VI. RECEIVER STRUCTURE FOR MULTIUSER INDOOR VLC SYSTEM

For multiuser indoor VLC systems, we used the NI-ADR shown in Fig. 9 [49]. This NI-ADR contains seven branches (1-7) and each branch has an array of four photo-detectors (2 × 2). Each photodetector has an area of 1 mm², which enable it to work at a data rate up to 4.4 Gb/s [15]. Due to the use of WDM, each photodetector in each branch of the NI-ADR was covered by a different optical bandpass filter. Four WDM channels are used; hence, four optical filters (red, yellow, green and blue) are utilized as shown in Fig. 9. Thus, each photodetector responds to a specific wavelength. The $El$ angle of the first face was set at 90°, whereas the other six branches were given an $El$ of 60°. The $A_z$ angles of the branches were set at 0°, 0°, 60°, 120°, 180°, 240° and 300°. The photodetector's FOV in each face was set to 20°. The values of $A_{zs}$, $El$s and FOVs of the NI-ADR were selected to make the NI-ADR view at least one TB at any receiver location on the room's communication floor. We used the SB scheme for the photo-detectors that have the same colour filters (see Fig. 9).

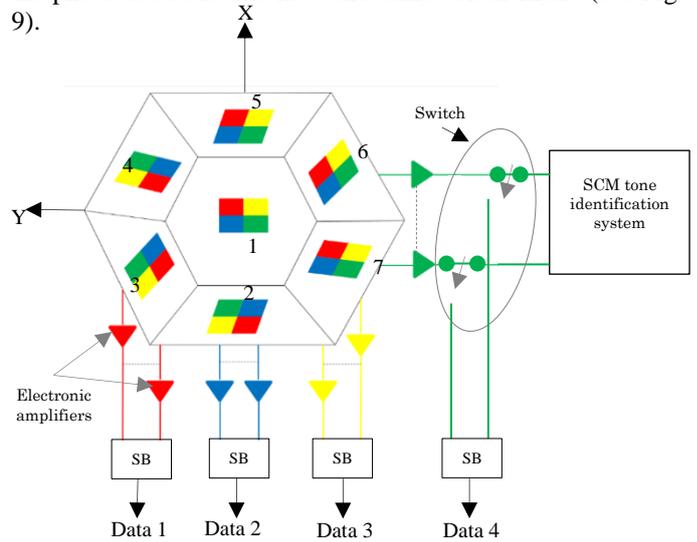

Fig. 9: NI-ADR configuration

As mentioned, the green channel is utilized to convey the SCM tones for setting up the transmission between users and transmitters, which means that data is not transmitted at the beginning of the connection. Therefore, the outputs of the green photodetectors of each user fed the SCM tone identification system to find the optimum TB for each user as can be seen in Fig. 9.

The SCM tone identification system is utilized to match each user with a TB that offers a good communication link without needing to know the location of the user. In addition, these SCM tones are utilized to obtain the CCI level at each user. As can be seen in Fig. 10, the SCM tones are used to calculate the output power of the photodetectors covered by the green colour at the beginning of the connection. As each TB is given a unique SCM frequency, electrical bandpass filters were used to separate these SCM tones (see Fig. 10). Each of these BPFs was given a centre frequency equal to the frequency of a SCM tone. The frequency range that was given to the SCM tones was selected close to DC where the indoor channel response has low attenuation. From Table II, it can be seen that the lowest 3dB channel bandwidth is 4.5 GHz. Thus, the frequency range chosen for SCM tones is 500 MHz to 3.8 GHz with 60 MHz guard. In addition, the bandwidth of the BPF was chosen to be 4 MHz. This decreases the total noise observed by the SCM tones and allows for SCM oscillator drift and BPF tolerances. The output of the electrical BPFs is a SCM tone plus noise.

The green colour of the RYGB LDs in each TB are used to carry the SCM tones at the beginning of the communication session to find the optimum TB for each user. Based on the location of the NI-ADR, more than one TB can be seen by the NI-ADR. Thus, to assess the ability of the SCM tone identification system to allocate each optical receiver to its closest TB, we determined two key distributions through simulations. Firstly, we obtained the distribution of the received electrical current due to the best TB ($C\_ds$). This is the desired SCM tone. Secondly, we obtained the distribution of the received electrical current due to the second best TB ($C\_uds$), which is the undesired SCM tone. Consequently, we calculated the probability of wrong decisions in the SCM tone identification system. We considered 1000 random positions of the NI-ADR on the communication floor to determine the distribution of $C\_ds$ and $C\_uds$. At each position of the NI-ADR, the values of $C\_ds$ and $C\_uds$ were calculated. Fig. 10 (a) and Fig. 10 (b) show the histograms and curve fittings of the $C\_ds$ and $C\_ds$ parameters respectively.

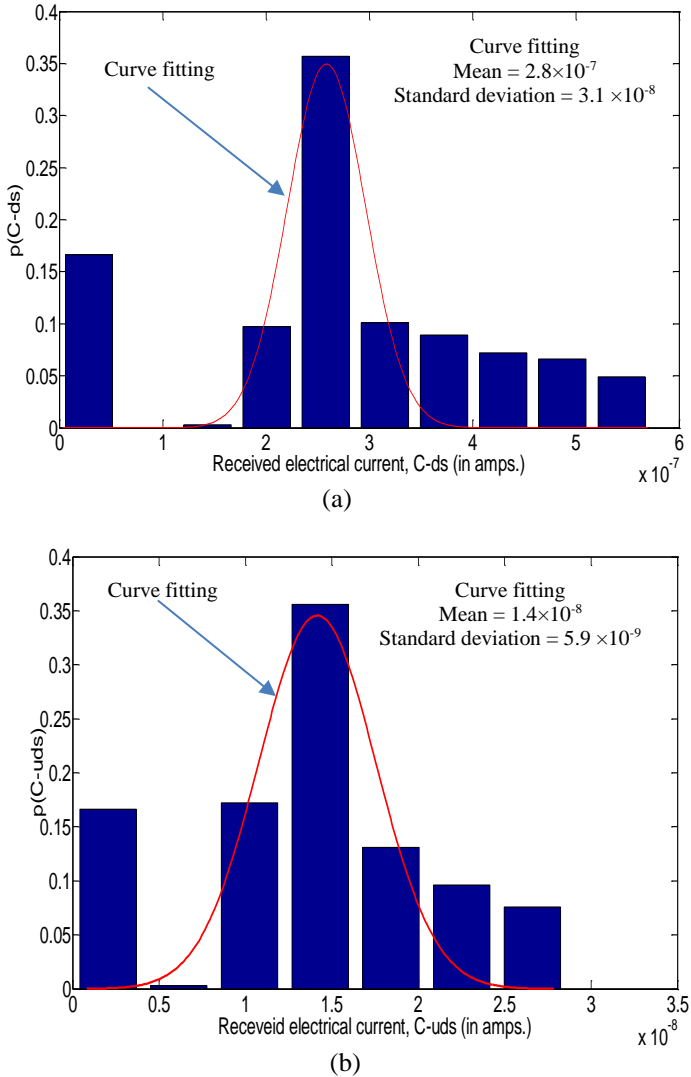

Fig. 10: Histogram and curve fitting of (a) $C\_ds$ and (b) $C\_uds$.

The normalized probability density functions (pdfs) of $C\_ds$ ($p(C\_ds)$) and $C\_uds$ ($p(C\_uds)$) can be written as:

$$p(C\_ds) = \frac{1}{\sqrt{2\pi}\,\sigma_{ds}} e^{-\left(\frac{C\_ds - m_{ds}}{\sqrt{2}\sigma_{ds}}\right)^2} \quad (19)$$

and

$$p(C\_uds) = \frac{1}{\sqrt{2\pi}\,\sigma_{us}} e^{-\left(\frac{C\_uds - m_{us}}{\sqrt{2}\sigma_{us}}\right)^2} \quad (20)$$

We assumed a Gaussian distribution for the received electrical current, which is reasonable given the multiple independent reflection surfaces and the observed results. Here, $\sigma_{ds}$ and $m_{ds}$ are the standard deviation and the mean value of $C\_ds$, respectively, and $\sigma_{us}$ and $m_{us}$ are the standard deviation and the mean value of $C\_uds$, respectively.

It should be noted that the output current ($z$) of each BPF is either the desired SCM tone (the SCM tone sent from the best TB) plus noise ($C\_ds + n$) or the undesired SCM tone (the SCM tone transmitted from the interfering TBs) plus noise ($C\_uds + n$); $n$ is the total noise seen by each SCM tone and is white Gaussian zero mean, with total standard deviation $\sigma_t$. It should be noted that for the SCM tones we used the bandwidth of the BPFs to calculate $\sigma_t$.

Following the analysis in [46], [49], we identify two hypotheses associated with $z$:
Hypothesis 1 ($H1$): $z = C\_uds + n$.
Hypothesis 2 ($H2$): $z = C\_ds + n$.
The pdfs of $z$ given $H1$ and $H2$ can be written as follows: under $H1$, $z$ is the convolution of the undesired SCM tone pdf and the noise pdf:

$$fz(z|H1) = p(C\_uds) \otimes p(n) \quad (21)$$

Solving equation (21), $fz(z|H1)$ can be written as:

$$fz(z|H1) = \frac{1}{\sqrt{2\pi(\sigma_{us}^2 + \sigma_t^2)}} e^{-\left(\frac{z - m_{us}}{\sqrt{2(\sigma_{us}^2 + \sigma_t^2)}}\right)^2} \quad (22)$$

Under $H2$, $z$ is the convolution of the desired SCM tone pdf and the noise pdf, which is given as:

$$fz(z|H2) = \frac{1}{\sqrt{2\pi(\sigma_{ds}^2 + \sigma_t^2)}} e^{-\left(\frac{z - m_{ds}}{\sqrt{2(\sigma_{ds}^2 + \sigma_t^2)}}\right)^2} \quad (23)$$

Applying a likelihood ratio to equations (22) and (23), we get:

$$\frac{fz(z|H2)}{fz(z|H1)} \underset{H1}{\overset{H2}{\gtrless}} 1$$

$$= \frac{\frac{1}{\sqrt{2\pi(\sigma_{ds}^2 + \sigma_t^2)}} e^{-\left(\frac{z - m_{ds}}{\sqrt{2(\sigma_{ds}^2 + \sigma_t^2)}}\right)^2}}{\frac{1}{\sqrt{2\pi(\sigma_{us}^2 + \sigma_t^2)}} e^{-\left(\frac{z - m_{us}}{\sqrt{2(\sigma_{us}^2 + \sigma_t^2)}}\right)^2}} \underset{H1}{\overset{H2}{\gtrless}} 1 \quad (24)$$

Solving equation (24), we get:

$$z \underset{H1}{\overset{H2}{\gtrless}} \frac{1}{\sigma_{ds}^2 - \sigma_{us}^2}\Bigg( m_{us}(\sigma_{ds}^2 + \sigma_t^2) - m_{ds}(\sigma_{us}^2 + \sigma_t^2) +$$

$$\sqrt{\begin{pmatrix} \sigma_{ds}^2\sigma_{us}^4 + \sigma_t^4\sigma_{us}^2 + \sigma_t^2\sigma_{us}^4 \\ -\sigma_{ds}^4\sigma_t^2 - \sigma_{ds}^4\sigma_{us}^2 - \sigma_{ds}^2\sigma_t^4 \end{pmatrix} \ln\left(\frac{\sigma_{us}^2 + \sigma_t^2}{\sigma_{ds}^2 + \sigma_t^2}\right) + \\ (m_{ds} - m_{us})^2(\sigma_t^2\sigma_{ds}^2 + \sigma_{ds}^2\sigma_{us}^2 + \sigma_t^4 + \sigma_t^2\sigma_{us}^2)} \Bigg) =$$

$$z \underset{H1}{\overset{H2}{\gtrless}} opt\_th \quad (25)$$

Thus, the probability of correct detection of the desired SCM tone ($Pcds$) is:

$$Pcds = \int_{opt\_th}^{\infty} fz(z|H2)dz$$
$$= \int_{opt\_th}^{\infty} \frac{1}{\sqrt{2\pi(\sigma_{ds}^2 + \sigma_t^2)}} e^{-\left(\frac{z-m_{ds}}{\sqrt{2(\sigma_{ds}^2+\sigma_t^2)}}\right)^2} dz \quad (26)$$

Consequently, the probability of correct detection of the undesired SCM tone ($Pfus$) is:

$$Pfus = \int_{opt_{th}}^{\infty} fz(z|H1)dz$$
$$= \int_{opt_{th}}^{\infty} \frac{1}{\sqrt{2\pi(\sigma_{us}^2 + \sigma_t^2)}} e^{-\left(\frac{z-m_{us}}{\sqrt{2(\sigma_{us}^2+\sigma_t^2)}}\right)^2} dz \quad (27)$$

Thus, the probability of not allocating the undesired SCM tone to a user ($Pcus$) is:

$$Pcus = 1 - Pfus \quad (28)$$

Thus, the overall probability of the SCM identification system making a correct dissection ($Pcd$) is:

$$Pcd = Pcds \, (Pcus)^{K-1} \quad (29)$$

where $K$ is the number of the TBs. Consequently, the probability of making the wrong decision in the SCM identification system is $Pwd$ as $1 - Pcd$. In our system, and for the given set of parameters, $Pwd$ is $8.1 \times 10^{-9}$. This value shows that the SCM tone identification system is able to find the optimum TB for each user with a high accuracy.

value of $CNR$ related with each SCM tone. Hence, the TBs are sorted in a descending order by the controller. It should be noted that each optical receiver has a different descending order of TBs beginning with the TB that gives the highest $CNR$ and ending with the TB that offers the lowest $CNR$. Therefore, the controller assigns to each user the TB that yields the highest $CNR$ from its group. For uplink transmission, we used the IR uplink design in [40]. In addition, each user was assigned a time slot to send the feedback information to the controller, which prevents interference in the uplink. The $CNR$ of any SCM tone at any optical receiver is written as [46], [49]:

$$CNR = \frac{(R_g \, Pr)^2}{2 \, \sigma_{ts}^2} \quad (30)$$

where $R_g$ is the photo-detector's responsivity for the green colour, $Pr$ is the optical power received by the green photo-detector and $\sigma_{ts}$ is the total noise standard deviation seen by each SCM tone. It should be noted that to calculate $\sigma_{ts}$, we considered the bandwidth of the BPF.

The SCM tones are also used to calculate the CCI level. However, no interference occurs between these SCM tones as shown in Fig. 11. Thus, we defined the CCI level at any user, as the total received power of all SCM tones except the one that was allocated to the user. For instance, if the allocated tone of the $mth$ TB is $f_m$, the level of CCI at the $nth$ user ($I_n$) because of the other SCM tones can be given as [46], [49]:

$$I_n = \sum_{\substack{k=1 \\ k \neq m}}^{Ma} \left(\frac{R_g \, Pr_{n,k}}{2}\right)^2, \quad n \in [1, 2 \dots N],$$
$$m \in [1, 2 \dots Ma] \quad (31)$$

where $Ma$ denotes the total number of active TBs (the TBs allocated to other optical receiver to transmit data) and $N$ denotes the number of optical receivers.

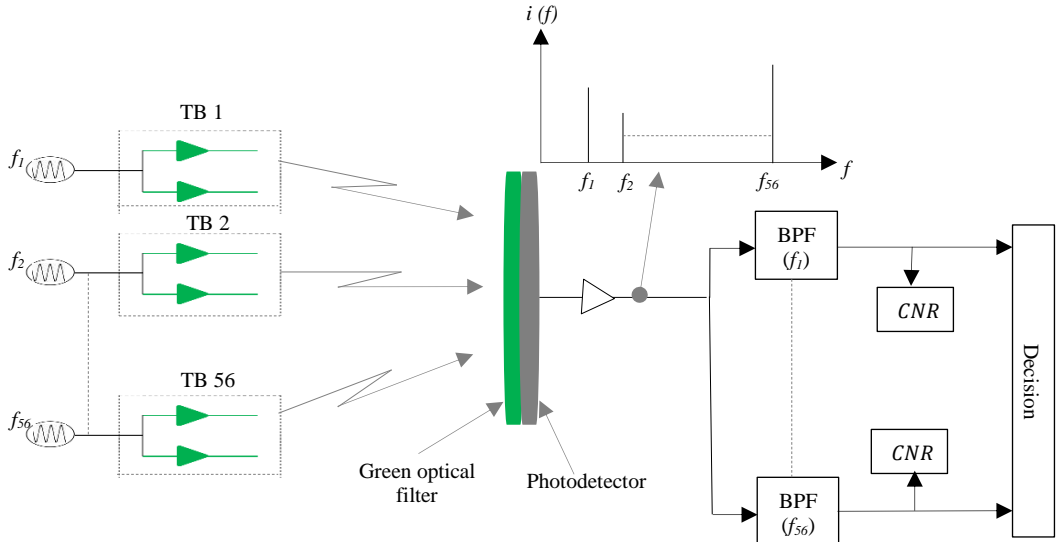

Fig. 11: Structure of SCM tone identification system.

As seen in Fig. 1, the controller is utilized to manage the connection between transmitters and users. The carrier to noise ($CNR$) ratio of the SCM tones is obtained at each optical receiver. Each optical receiver notifies the controller of the

## VII. PERFORMANCE ANALYSIS OF DATA CHANNELS

Once the controller assigns a TB to each user, the data is transmitted through the four channels of the RYGB LDs. In this system, we consider the effect of CCI interference. Thus, the signal to interference to noise ($SINR$) ratio is utilized to assess the performance of this system. In general, the $SINR$ of the *nth* user is given as [29], [50]:

$$SINR_n = \frac{R^2(P_{s1} - P_{s0})_n^{\,2}}{\sigma_t^{\,2} + I_n} \qquad (32)$$

It should be noted that we used the green channel to estimate the level of the CCI. For each user, we obtained the CCI level of other channels from the green channel. Each channel of the RYGB LD has a different transmitted optical power (to obtain an acceptable white colour [26]). In addition, each photodetector in each branch of the NI-ADR sees the same room area and each photodetector has a specific optical filter that has a different responsivity. Thus, the CCI level of the data channels of the *nth* user ($I_n$) is calculated from the CCI level obtained from SCM tones as [49]:

$$I_{red} = \left(\frac{R_r}{R_g}\right)^2 \left(\frac{Pt_r}{Pt_g}\right) I_{green}, I_{yellow} \left(\frac{R_y}{R_g}\right)^2 \left(\frac{Pt_y}{Pt_g}\right) I_{green}$$

$$\text{and } I_{blue} \left(\frac{R_b}{R_g}\right)^2 \left(\frac{Pt_b}{Pt_g}\right) I_{green} \qquad (33)$$

where $R_r$, $R_y$ and $R_b$ are the responsivities of the red photodetector, the yellow photodetector and the blue photodetector, respectively and $Pt_r$, $Pt_g$, $Pt_y$ and $Pt_b$ are the red, green, yellow and blue optical transmitted powers, respectively. Hence, the $SINR$ of each data channel at any user is given as [49]:

$$SINR_{red} = \frac{R_r^2 (P_{s1} - P_{s0})^2}{\sigma_t^{\,2} + \left(\frac{R_r}{R_g}\right)^2 \left(\frac{Pt_r}{Pt_g}\right) I_{green}} \qquad (34)$$

$$SINR_{yellow} = \frac{R_y^2 (P_{s1} - P_{s0})^2}{\sigma_t^{\,2} + \left(\frac{R_y}{R_g}\right)^2 \left(\frac{Pt_y}{Pt_g}\right) I_{green}} \qquad (35)$$

$$SINR_{blue} = \frac{R_b^2 (P_{s1} - P_{s0})^2}{\sigma_t^{\,2} + \left(\frac{R_b}{R_g}\right)^2 \left(\frac{Pt_b}{Pt_g}\right) I_{green}} \qquad (36)$$

$$SINR_{green} = \frac{R_g^2 (P_{s1} - P_{s0})^2}{\sigma_t^{\,2} + I_{green}} \qquad (37)$$

Table III illustrates the simulation parameters that are utilized in the multiuser VLC system.

TABLE III
SIMULATION PARAMETERS OF MULTIUSER VLC SYSTEM

| Parameters | Configurations |
|---|---|
| Red LD optical power | 0.8W |
| Yellow LD optical power | 0.5W |
| Green LD optical power | 0.3W |
| Blue LD optical power | 0.3W |
| Number of ADR branches | 7 |
| Number of photodetectors/branch | 4 |
| Photodetector's FOV | 20º |
| Elevation of each branch | 90º, 60º, 60º, 60º, 60º, 60º, 60º |
| Azimuth of each branch | 0º, 0º, 60º, 120º, 180º, 240º 300º |
| Photodetector's area | 1 mm² |
| Red photodetector's responsivity | 0.4 |
| Yellow photodetector's responsivity | 0.35 |
| Green photodetector's responsivity | 0.3 |
| Blue photodetector's responsivity | 0.2 |

Each TB can serve up to four users simultaneously. Therefore, to assess the performance of the multiuser VLC system, we determined the maximum data rate that can be transmitted by each channel of the user located at (0.5 m, 0.5 m, 1 m) versus an increase in the number of active users. We considered each channel carries the maximum data rate that results in BER not exceeding $10^{-6}$, which gives a reliable connection between users and transmitters. We obtained the maximum data rate for a user located at (0.5 m, 0.5 m, 1 m,) as this location represents the worst case. This is attributed to reflections that are high at this location (high diffuse reflections and high CCI). Hence, this ensured that the other users have better or equal performance to this user. In this system, we considered the effect of diffuse reflections and CCI to obtain the performance of the system. It is worth mentioning that each channel has a 3dB channel bandwidth (22.3 GHz) similar to that mentioned in Table II at (0.5 m, 0.5 m, 1 m). CCI occurs when signals from the interfering TBs are received by the optical receiver.

The interference happens due to either LOS components and/or reflection components. However, we used NI-ADR, which has many faces pointed to different locations. Thus, the controller can assign two TBs from two different directions when two users are located at the same location as shown in Fig. 12. Hence, our system guarantees that there is no CCI due to LOS components (just due to reflection components). In addition, when the number of optical receivers is large, the controller can allocate one or two channel(s) to each user as the TB can serve up to four users at the same time.

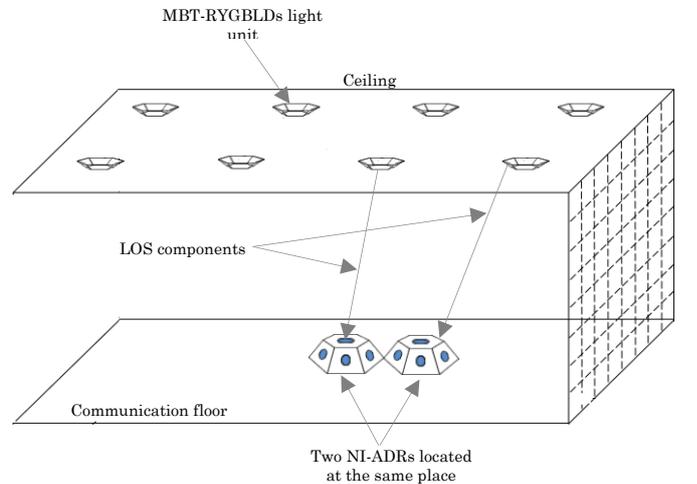

Fig. 12: Two NI-ADRs located at same place and served by different directions of TBs to prevent interference due to LOS components.

Fig. 13 depicts the impact of an increase in the number of optical receivers on each channel data rate of the optical receiver located at (0.5 m, 0.5 m 1 m). In addition, the aggregate data rate per optical receiver when this optical receiver was allocated four channels is also shown in Fig. 13. The data rate of each colour was calculated when the BER is $10^{-6}$ (SINR = 13.6 dB).

As can be seen in Fig. 13, the red channel offers a higher data rate than the other channels. This is due to the higher transmitted optical power of the red LD compared to the other LDs (see Table III). In addition, the responsivity of the red photodetector is higher than the responsivity of the yellow, green and blue photodetectors. Due to the use of MBT as a transmitter and NI-ADR as a receiver, this leads to eliminate the interference due to LOS components (CCI happens due to reflection components only). In addition, the limited FOV (FOV = 20º) of each photodetector in each face of the NI-ADR, limited the range of the rays captured by each photodetector and reduced the effect of CCI due to the reflection components. Thus, it can be seen that an increase in the number of optical receivers leads to a slight decrease in the data rate of each colour at BER of $10^{-6}$.

When each TB serves one optical receiver (i.e. the four channels of the TB are assigned to one optical receiver), the VLC system can support 56 optical receivers with an aggregate data rate per optical receiver not less than 16.3 Gb/s as shown in Fig. 13. However, each TB can serve up to four devices simultaneously (when these devices are located inside the coverage area of the TB). Therefore, when each device is given only one channel, our proposed VLC system can support up to 224 devices at a data rate not less than 2.15 Gb/s (the minimum data rate of the blue channel) as can be seen in Fig. 13. We do not consider the fairness between devices in this work. Thus, the controller assigns one channel for each device in a random way in this case. The aggregate capacity of our proposed VLC systems when fully loaded can be calculated based on the number of the MBT-RYGB LDs light units in the room (eight), the number of TBs per MBT-RYGB LDs light units (seven), the number of colours per TB (four) and the data rate achieved by each colour. Thus, our proposed VLC system can achieve an aggregate data rate of 912.8 Gb/s. Hence, our suggested system may be deployed for indoor internet of things (IoT) applications as many devices (with a high data rate) can be served in a small area.

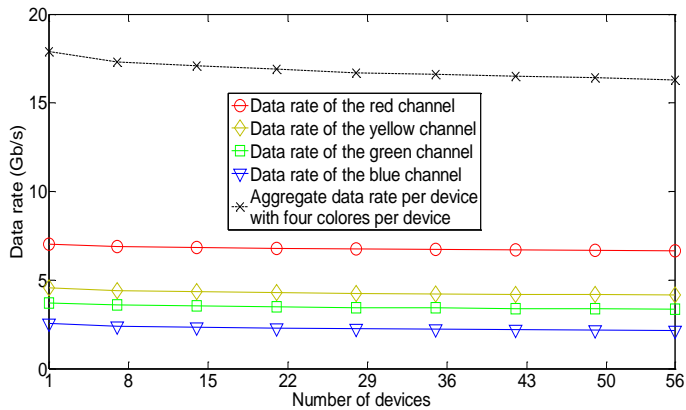

Fig. 13: Data rate of each channel and aggregate data rate per optical receiver placed at (0.5 m, 0.5 m, 1 m) versus number of optical receivers.

## VIII. IMPACT OF THE USERS MOBILITY ON THE SINR.

In this section we consider the effect of the users' mobility on the performance of the multi-user VLC system by calculating the SINR at many locations in the room and obtaining the cumulative distribution function (CDF) of the SINR. In addition, to take into account the probability that a user may be at any given location in the room and to determine the probability that a given SINR is observed, we used a Markov chain to model users' movement where discrete locations in the room become states in the Markov chain and users' movement is then a set of transitions between these states.

Fig. 14 illustrates the CDF of the SINR of the MBT-VLC system when the system operates at an aggregate data rate of 15 Gb/s and the NI-ADR moves along $x = 0.5$ m, $x = 1$ m, $x = 1.5$ m, $x = 2$ m and when the NI-ADR was randomly located (1000 random locations) in the communication floor of the room. In addition, Fig. 14 shows the Results of aggregating all the data for all lines ($x = 0.5$ m, $x = 1$ m, $x = 1.5$ m, $x = 2$ m). It should be noted that due to the symmetry of the room, we obtained the results when the user moves along the *y*-axis and at $x = 0.5$ m, $x = 1$ m, $x = 1.5$ m, $x = 2$ m in steps of 0.5 m. As can be seen in Fig. 14, the performance of the system was better when the mobile receiver moved along the $x = 1$ m. This is because of the optical receiver proximity to light units along the x = 1 m when compared with $x = 0.5$ m, $x = 1.5$ m and $x = 2$ m. In addition, the CDF of the SINR is shown when the NI-ADR was randomly distributed (1000 locations) on the communication floor of the room. This results is comparable with the CDF of the SINR that results from aggregating all data for all the lines ($x = 0.5$ m, $x = 1$ m, $x = 1.5$ m, $x = 2$ m).

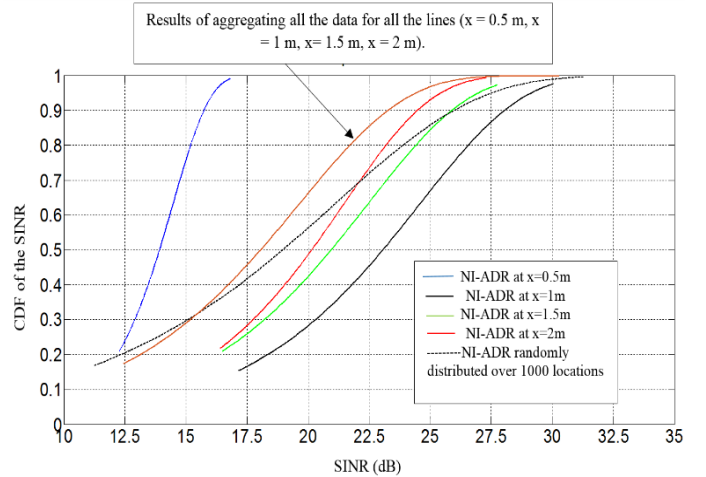

Fig. 14: CDF of the SINR of the MBT-VLC system when the system operates at 15 Gb/s and the NI-ADR moves along $x = 0.5$ m, $x = 1$ m, $x = 1.5$ m, $x = 2$ m and when the NI-ADR was randomly located (1000 random locations) in the room.

Although the analysis conducted in Fig. 14 determined the SINR at many locations in the room, yet it does not take into account the probability that a user may be at this location, hence does not determine the probability that a given SINR is observed. To take the movement of the users into account, we have introduced a Markov chain analysis to find the CDF of the SINR while considering the probability that a user may be at a given location, hence we determined the probability that a given SINR is observed, based on people movement. This is the first time people movement and Markov models of such movement are used in multiuser indoor VLC systems, to the best of our knowledge. Thus, we modelled the location of the user in the room at a given time as a state in the Markov Chain. To obtain the probability of the user occupying a given location/state we used an *M/M/1/N* queuing model, where each line (for example the line *x* = 1 m) in the room is modelled separately and acts as a single server that has *N* locations that can be occupied by users. We assumed that λ is the arrival rate of users into the room and μ is the users' departure rate out of the room. In addition, we considered movement such that a user moves in a steps of 0.5 m along the y-axis and at *x* = 0.5 m, *x* = 1 m, *x* = 1.5 m or *x* = 2 m, which means *N* = 14 states in the *M/M/1/N* model.

When the arrival rate (λ) of the users into the room is much smaller than the departures rate (μ) of the users from the room (i.e. λ<< μ), the queue is almost empty most of the time, which means that users are near to the room's entrance most of the time. When λ increases, the queue increases in size and this models people moving into the room. In addition, when λ is large and the queue states are all occupied, this models a room full of people.

The Markov chain determines *p(k)*, which is the probability that the queue has *k* people and thus, the probability of each state/location in the room. Therefore, each SINR observed in the room at a given location is now attached to a probability (probability of occupation) of the space. The probability *p(k)* is given as [51]:

$$p(k) = \frac{\rho^k(1-\rho)}{1-\rho^{N+1}} \quad (38)$$

where $\rho = \frac{\lambda}{\mu}, \rho \neq 1$. It should be noted that when $\rho = 1$, $p(k) = \frac{1}{N+1}$ and in this case all locations in the room are occupied and all locations in the room have the same probability of occupancy as was illustrated in Fig. 14.

Fig. 15 illustrates the CDF of the SINR of the MBT-VLC system when the system operates at an aggregate data rate of 15 Gb/s and the NI-ADR moves along *x* = 0.5 m, *x* = 1 m, *x* = 1.5 m, *x* = 2 m. In addition, we considered the effects of the arrival rate and departure rate of users. The results were obtained when $\rho = 0.3$, $\rho = 0.8$ and $\rho = 0.9$ and were compared with the simulation results obtained in Fig. 14. Reducing the value of $\rho$ leads to users occupying locations near the entrance of the room. These are locations that have lower SINR. Therefore, reducing the value of $\rho$ leads to a reduction in the possible locations that offer a good connection. It should be noted that we can "modulate" the CDF of the SINR by the users' arrival rates and users' departure rates.

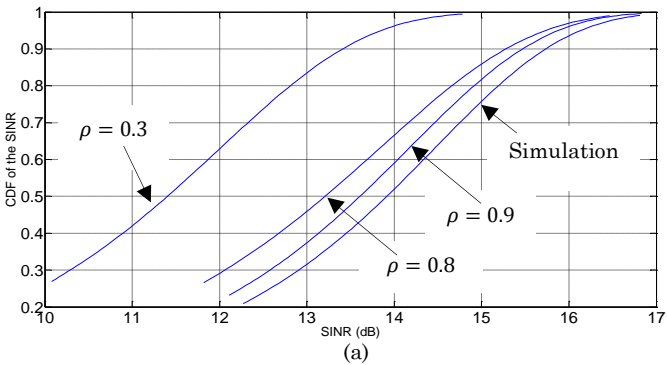

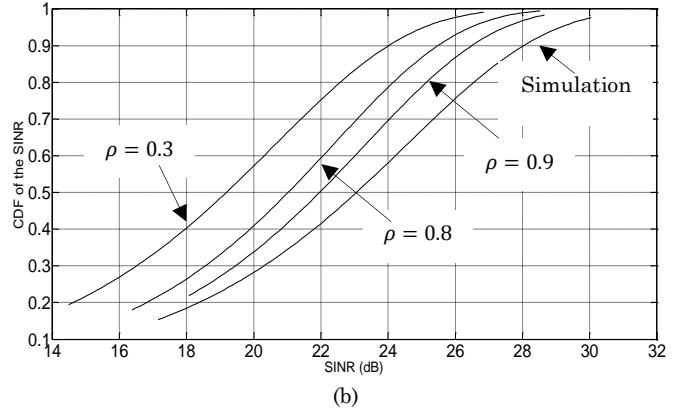

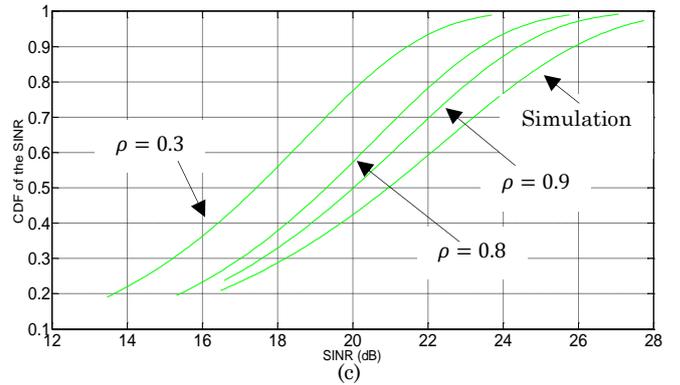

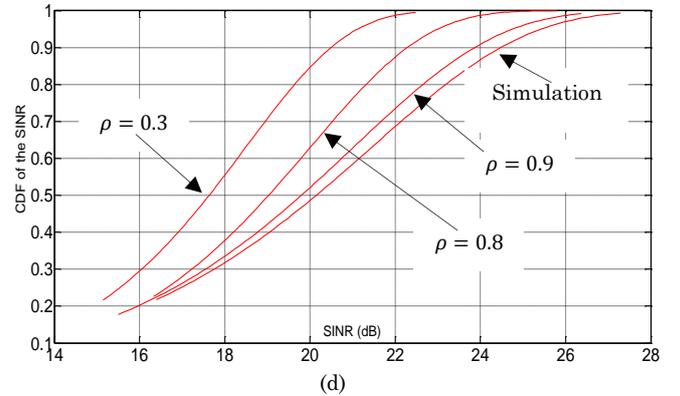

Fig. 15: CDF of the SINR of the MBT-VLC system at different values of $\rho$ when the system operates at 15 Gb/s and the NI-ADR moves along: (a) *x* = 0.5 m, (b) *x* = 1 m, (c) *x* = 1.5 m and (d) *x* = 2 m.

## IX. CONCLUSIONS.

In this work, we used a MBT to improve the performance of indoor VLC systems. The MBT had seven TBs and each one of these TBs is directed to a specific location on the room's communication floor. This led to an improvement in the 3dB channel bandwidth of the indoor VLC system and therefore an increase in the received optical power. Two VLC systems were proposed based on MBT: a single user VLC system and a multiuser VLC system. For the single user VLC system, we used a W-FOV receiver and an ADR as optical receivers. The results showed that the single user VLC system offers a data rate of 4 Gb/s and 10 G b/s when using W-FOV receiver and ADR, respectively. The influence of diffuse reflections (up to second order), mobility and lighting constraint were taken into account for the single user VLC system.

In the multiuser VLC system, we proposed SCM tones for resource allocation. Thus, each optical receiver was assigned a TB that offers good performance without the need to know the location of the optical receiver. RYGB LDs were used as luminaires. We, therefore, used WDM to achieve a higher data rate for each optical receiver. In the multiuser VLC system, we proposed NI-ADR as an optical receiver where each photodetector was covered by a specific optical bandpass filter. We considered the effect of the CCI between transmitters in the multiuser VLC system. The results showed that this system can support up to 56 devices when each device was allocated four channels at a data rate not less than 16.3 Gb/s and BER not exceeded $10^{-6}$. When each user was allocated one channel, this enabled the TB to serve up to four users simultaneously. Hence, the system can support 224 devices at a data rate not less than 2.15 Gb/s. Therefore, this system may be deployed for indoor IoT applications as many devices (with a high data rate) can be served in a small area. In addition, we used Markov chain to modulate the users' mobility for multi-user indoor VLC system.

ACKNOWLEDGEMENTS

Safwan Hafeedh Younus would like to thank Ninevah University for financial support and funding his PhD scholarship. This work was supported by the Engineering and Physical Sciences Research Council (ESPRC), INTERNET (EP/H040536/1), STAR (EP/K016873/1) and TOWS (EP/S016570/1) projects. All data are provided in full in the results section of this paper.